\documentclass[showpacs,preprintnumbers,twocolumn,
amsmath,amssymb,groupedaddress,superscriptaddress]{revtex4}
\usepackage{graphicx}
\usepackage{dcolumn}
\usepackage{bm}

\begin{document}

\preprint{}

\title{Superscaling in Dilute Fermi Gas and Its Relation to General Properties
of\\ the Nucleon Momentum Distribution in Nuclei}

\author{A.N. Antonov}
\affiliation{Institute for Nuclear Research and Nuclear Energy,
Bulgarian Academy of Sciences, Sofia 1784, Bulgaria}

\author{M.V. Ivanov}
\affiliation{Institute for Nuclear Research and Nuclear Energy,
Bulgarian Academy of Sciences, Sofia 1784, Bulgaria}

\author{M.K. Gaidarov}
\affiliation{Institute for Nuclear Research and Nuclear Energy,
Bulgarian Academy of Sciences, Sofia 1784, Bulgaria}

\affiliation{Instituto de Estructura de la Materia, CSIC, Serrano
123, E-28006 Madrid, Spain}

\author{E. Moya de Guerra}
\affiliation{Instituto de Estructura de la Materia, CSIC, Serrano
123, E-28006 Madrid, Spain}

\affiliation{Departamento de Fisica Atomica, Molecular y Nuclear,
Facultad de Ciencias Fisicas, Universidad Complutense de Madrid,
E-28040 Madrid, Spain}


\begin{abstract}

The superscaling observed in inclusive electron scattering is
described within the dilute Fermi gas model with interaction
between the particles. The comparison with the relativistic Fermi
gas (RFG) model without interaction shows an improvement in the
explanation of the scaling function $f(\psi ')$ in the region
$\psi ' < -1$, where the RFG result is $f(\psi ') = 0$. It is
found that the behavior of $f(\psi ')$ for $\psi ' < -1$ depends
on the particular form of the general power-law asymptotics of the
momentum distribution $n(k)\sim  1/ k^{4+m}$ at large $k$. The
best agreement with the empirical scaling function is found for
$m\simeq  4.5$ in agreement with the asymptotics of $n(k)$ in the
coherent density fluctuation model where $m = 4$. Thus,
superscaling gives information about the asymptotics of $n(k)$ and
the NN forces.

\end{abstract}


\maketitle

\section[]{INTRODUCTION\label{sect1ant}}

The concepts of
$y$-scaling~\cite{West75,Sick80,Cio87,Day90,Cio90,Cio96,CW97,CW99,FCW00,AMD+88}
and superscaling (based on $\psi'$-scaling
variable)~\cite{AMD+88,BCD+98,DS99l,DS99,MDS02,AGK+04,AGK+05,AIG+06,AIG+06a}
have been used for extensive analyses of the vast amount of
inclusive electron scattering world data (see
also~\cite{Benhar2006}). These analyses showed the existence of
high-momentum components in the nucleon momentum distribution (MD)
$n(k)$ at momenta $k > 2$~fm$^{-1}$ due to the presence of
nucleon-nucleon (NN) correlations. Scaling of the first kind
(\emph{i.e.} no dependence on the momentum transfer) can be
observed at excitation energies below the quasielastic (QE) peak.
Scaling of second kind (\emph{i.e.} no dependence on the mass
number) is excellent in the same region. When scaling of first and
second type occur one says that superscaling takes place. It was
shown (\emph{e.g.}, in~\cite{AIG+06a,AIG+06,AGK+05,AGK+04}) that
the physical reason for the superscaling phenomena is the specific
high-momentum tail of $n(k)$ which is similar for all nuclei.

It has been shown~\cite{BCD+04,AlvRuso03} that due to the
contribution introduced by inelastic scattering together with the
correlation contribution and meson exchange
currents~\cite{Amaro2002,Pace03} both scaling of the first and, to
a lesser extent, of the second kind are violated at energies above
the QE peak.

The theoretical concept of superscaling has been introduced
in~\cite{AMD+88,BCD+98} using the properties of the relativistic
Fermi gas (RFG) model . The Fermi momentum for the RFG was used as
a physical scale to define the proper scaling variable $\psi '$
for each nucleus. As emphasized in~\cite{DS99}, however, the
actual dynamical physics content of the superscaling phenomenon is
more complex than that provided by the RFG model. In particular,
the extension of the superscaling property to large negative
values of $\psi '$ ($\psi ' < -1$) is not predicted by the RFG
model. The QE scaling function in the RFG model
$f_\text{RFG}^\text{QE}(\psi ') = 0$ for $\psi ' \leq -1$, whereas
the experimental scaling function $f^\text{QE}(\psi ')$ has been
observed for large negative values of $\psi '$ up to $\psi '
\approx  -2$ in the data for $(e,e')$ processes. Thus, it has been
necessary to consider the superscaling in theoretical approaches
which go beyond the RFG model. One of them is the coherent density
fluctuation model (CDFM) (\emph{e.g.},~\cite{ANP79+,AHP88}) which
gives a natural extension of the Fermi gas case to realistic
finite nuclear systems. It was shown
in~\cite{AIG+06a,AIG+06,AGK+05,AGK+04} that in the CDFM both basic
quantities , density and momentum distributions, are responsible
for the scaling and superscaling phenomena in nuclei. The results
of the CDFM for the QE scaling function $f(\psi ')$ agree with the
available experimental data at different transferred momenta and
energies below the QE peak position, showing superscaling for
$\psi '< 0$, including $\psi '\lesssim -1$ and going well beyond
the RFG model. Secondly, as pointed out in~\cite{AGK+05,AIG+06a},
the nucleon momentum distribution for various nuclei obtained
in~\cite{AGI+02} (and with the modification in~\cite{AIG+06a})
within a theoretical approach based on the light-front dynamics
(LFD) method (\emph{e.g.},~\cite{CK95,CDK98}) can also be used to
describe both $y$- and $\psi '$-scaling data.

The superscaling analyses of inclusive electron scattering from
nuclei (for energies of several hundred MeV to a few GeV) have
been extended in~\cite{Amaro2005} to include not only QE processes
but also the region where $\Delta $ excitation dominates. Both QE
scaling functions $f^\text{QE}(\psi ')$ and $f^{\Delta}(\psi
'_\Delta )$ in the delta region were deduced from phenomenological
fits to electron-nuclei scattering data. Generally, the specific
features of the scaling function should be accounted for by
reliable microscopic calculations that take final-state
interactions into account. In particular, the scaling function
$f^\text{QE}(\psi')$ with asymmetric shape obtained
in~\cite{Caballero2005,Caballero2006} by using a relativistic mean
field (RMF) for the final states agrees well with the experimental
scaling function.

The features of the superscaling phenomenon in inelastic electron
scattering have induced studies of neutrino scattering from nuclei
on the same basis. The neutrino-nucleus interactions have been
studied using the superscaling analyses of few-GeV inclusive
electron scattering data in a method proposed in~\cite{Amaro2005}
to predict the inclusive $\nu A$ and $\overline{\nu} A$ cross
sections for the case of $^{12}$C in the nuclear resonance region.
Various other theoretical considerations
(\emph{e.g.},~\cite{Meucci20041,Amaro2006,Barbaro2006,Martinez2006,Nieves2005,Barbaro2005,Maieron2003,Meucci20042,Benhar,Co,Leither,Sato})
have been devoted to studies of both neutral-
(\emph{e.g.},~\cite{Meucci20041,Amaro2006,Barbaro2006,Martinez2006,Nieves2005})
and charge-changing
(\emph{e.g.},~\cite{Amaro2005,Caballero2005,Barbaro2006,Martinez2006,Barbaro2005,Maieron2003,Meucci20042,Benhar,Co,Leither,Sato})
neutrino-nucleus scattering.

The CDFM and the LFD method have been extended in~\cite{AIG+06a}
from the QE to the delta-excitation region of the inclusive electron
scattering and the QE scaling functions calculated in both methods
have been used to calculate and to predict charge-changing
neutrino-nucleus cross sections of the $(\nu_\mu, \mu^-)$ and
$(\overline{\nu}_\mu, \mu^+)$ reactions on $^{12}$C at energies from
1 to 2~GeV. The asymmetry in the CDFM QE scaling function has been
introduced in a phenomenological way. These analyses make it
possible to gain information about the nucleon correlation effects
on both nucleon momentum and local density distributions. It became
clear that only the detailed knowledge of the behavior of $n(k)$ at
high momenta in realistic nuclear systems could lead to quantitative
agreement with the experimental scaling function. On the other hand,
the behavior of the latter gives a valuable information about the NN
correlation effects on the tail of the momentum distribution. So, it
was shown in~\cite{AGK+05} within the CDFM that the $y$- and $\psi
'$-scaling data are informative for $n(k)$ at momenta up to $k
\approx 2$--$2.5$~fm$^{-1}$ and it was concluded that further
experiments are necessary in studies of the high-momentum components
of the nucleon momentum distributions.

The aim of the present work is to consider in more detail the
connection between the NN forces in nuclear medium and their
effect on the components of $n(k)$ from one side and, from the
other side, the role of $n(k)$ on the behavior of the QE scaling
function. For this purpose we use firstly the MD in a hard-sphere
dilute Fermi gas model (HSDFG)
(\emph{e.g.},~\cite{Belyakov,Galitskii,Migdal,Czyz,Sartor}) to
calculate the scaling function. Secondly, we make an attempt to
throw light on the connection between the generally established
high-momentum asymptotics of
$n(k)$~\cite{Amado77,Amado76b,Amado76a,Czyz,Sartor} and the QE
scaling function. The latter makes it possible to establish (at
least approximately) the particular form of the power-law decrease
of $n(k)$ at large values of $k$. This makes it possible to
extract additional information about the NN forces from the
description of the superscaling phenomenon.

The theoretical scheme and the results of calculations are given in
Section~\ref{sect2ant}. The conclusions are summarized in
Section~\ref{sect3ant}.

\section[]{THEORETICAL SCHEME AND RESULTS OF CALCULATIONS\label{sect2ant}}

In the first part of this Section we consider the hard-sphere
dilute Fermi gas, \emph{i.e.} the low-density Fermi gas whose
particles interact via a repulsive hard-core potential (see,
\emph{e.g.},~\cite{Belyakov,Galitskii,Migdal,Czyz,Sartor}), and
use the nucleon momentum distribution in such a system to
calculate the scaling function $f^\text{HSDFG}(\psi ')$. The
quantities of interest in the HSDFG model can be expanded in
powers of the parameter $k_F c$, where $c$ ($>0$) denotes the
hard-core radius in the NN interactions or it is identified with
the scattering length in free space, and $k_F$ is the Fermi
momentum. In~\cite{Sartor} the value of $k_F c$ is adopted to be
equal to $0.70$ which corresponds to NN core radius of $c=
0.50$~fm and the typical value of the Fermi momentum $k_F =
1.40$~fm$^{-1}$. As was pointed out by Migdal~\cite{Migdal},
$n(k)$ in the normal Fermi gas is discontinuous at the Fermi
momentum. The analytical expressions for the dimensionless $n(k)$
in the HSDFG obtained in~\cite{Belyakov,Sartor} have the form:
\begin{equation}\label{eq1ant}
n(k)\!=\!n_{<}(k)\!+\!n_{>}(k)\text{ with }
\begin{cases}
n_{<}(k)\!=\!0\text{ for }k\!>\!k_F\\
n_{>}(k)\!=\!0\text{ for }k\!<\!k_F
\end{cases}\!\!\!\!.
\end{equation}

At $k<k_F$:
\begin{align}\label{eq2ant}
n_{<}(k)\!=&1\!-\!\frac{\nu\!-\!1}{3\pi^2
x}(k_Fc)^2\bigg[(7\ln2\!-\!8)x^3+(10\!-\!3\ln2)x\notag\\
\!+&2\ln\frac{1\!+\!x}{1\!-\!x}\!-\!2(2\!-\!x^2)^{3/2}\ln\frac{(2\!-\!x^2)^{1/2}\!+\!x}{(2\!-\!x^2)^{1/2}\!-\!x}\bigg],
\end{align}
where $x = k/k_F$ and $\nu=4$~\cite{Sartor} is adopted.

At $1<x<\sqrt{2}$:
\begin{align}\label{eq3ant}
n_{>}(k)\!=&\frac{\nu\!-\!1}{6\pi^2
x}(k_Fc)^2\bigg\{(7x^3\!-3x\!-\!6)\ln\frac{x\!-\!1}{x\!+\!1}\notag\\
+&(7x^3\!-3x\!+\!2)\ln2\!-\!8x^3\!+\!22x^2\!+\!6x\!-\!24\notag\\
+&2(2\!-\!x^2)^{3/2}\bigg[\ln\frac{2\!+\!x\!+\!(2\!-\!x^2)^{1/2}}{2\!+\!x\!-\!(2\!-\!x^2)^{1/2}}\notag\\
+&\ln\frac{1\!+\!(2\!-\!x^2)^{1/2}}{1\!-\!(2\!-\!x^2)^{1/2}}-2\ln\frac{x\!+\!(2\!-\!x^2)^{1/2}}{x\!-\!(2\!-\!x^2)^{1/2}}\bigg]\bigg\}.
\end{align}

At $\sqrt{2}<x<3$:
\begin{align}\label{eq4ant}
n_{>}(k)\!=&\frac{\nu\!-\!1}{6\pi^2
x}(k_Fc)^2\bigg\{(7x^3\!-3x\!-\!6)\ln\frac{x\!-\!1}{x\!+\!1}-\!8x^3\!+\!22x^2\notag\\
+&6x\!-\!24+(7x^3\!-3x\!+\!2)\ln2\!-\!4(x^2\!-\!2)^{3/2}\notag\\
\times&\bigg[\arctan\frac{(x\!+\!2)}{(x^2\!-\!2)^{1/2}}\!+\!\arctan\frac{1}{(x^2\!-\!2)^{1/2}}\notag\\
&~~-2\arctan\frac{x}{(x^2\!-\!2)^{1/2}}\bigg]\bigg\}.
\end{align}

At $x > 3$:
\begin{align}\label{eq5ant}
n_{>}(k)\!=&2\frac{\nu\!-\!1}{3\pi^2
x}(k_Fc)^2\bigg\{2\ln\frac{x\!+\!1}{x\!-\!1}-\!2x\!+\!(x^2-2)^{3/2}\notag\\
\times&\bigg[2\arctan\frac{x}{(x^2\!-\!2)^{1/2}}\!-\!\arctan\frac{x-2}{(x^2\!-\!2)^{1/2}}\notag\\
&~~-\arctan\frac{(x+2)}{(x^2\!-\!2)^{1/2}}\bigg]\bigg\}.
\end{align}

\begin{figure}
\centering
\includegraphics[width=80mm]{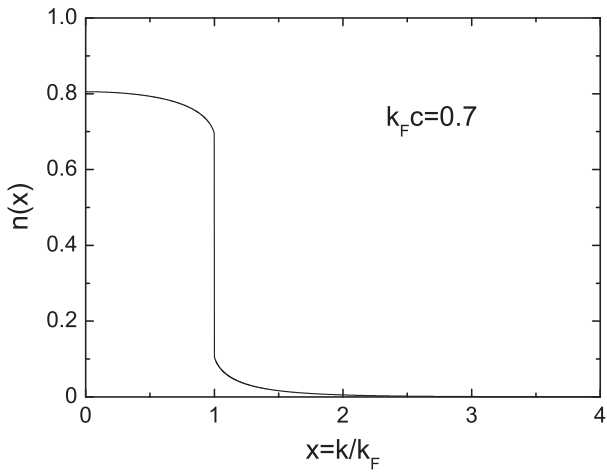}
\caption{\label{fig1ant} The momentum distribution $n(k)$ in
hard-sphere dilute Fermi gas~\cite{Sartor} as a function of $x =
k/k_F$.}
\end{figure}

The momentum distribution in the HSDFG model is presented in
Fig.~\ref{fig1ant} for the value of $k_Fc=0.70$.

Following the definition of the $\psi' $-scaling function given by
Barbaro \emph{et al.}~\cite{BCD+98}, one can obtain for the case
of the HSDFG:
\begin{equation}\label{eq6ant}
f^\text{HSDFG}(\psi')=\frac{3}{2}\int\limits_{|\zeta|/\eta_F}^{\infty}
x n(x)dx,
\end{equation}
where $\eta _F = k_F/ m_N$, $m_N$ being the nucleon mass, and
\begin{equation}\label{eq7ant}
\zeta=\psi'\Big\{\Big[\sqrt{1+\eta_F^2}-1\Big]\Big[2+\psi'^2\Big(\sqrt{1+\eta_F^2}-1\Big)\Big]\Big\}^{1/2}.
\end{equation}

\begin{figure}
\centering
\includegraphics[width=80mm]{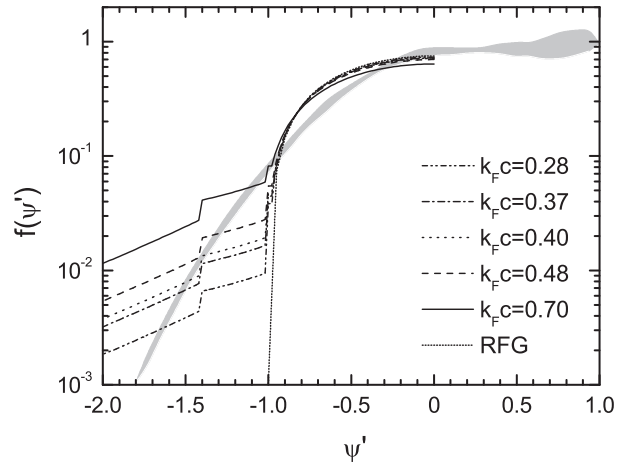}
\caption{\label{fig2ant} The scaling function $f(\psi ')$ in HSDFG
calculated for different values of $k_Fc$ in comparison with the RFG
model result. The grey area shows experimental data taken
from~\cite{DS99}.}
\end{figure}

\begin{figure}
\centering
\includegraphics[width=80mm]{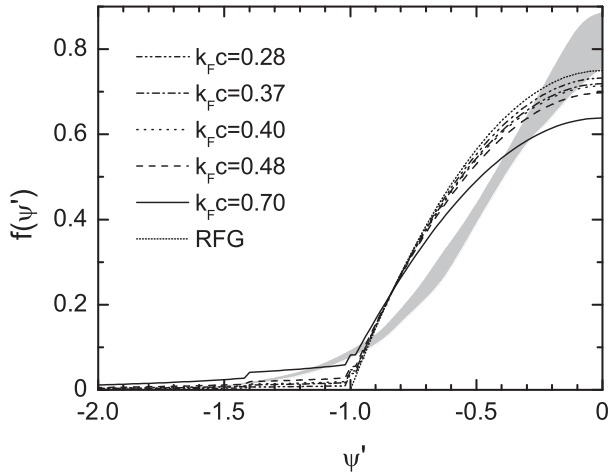}
\caption{\label{fig3ant} The same as in Fig.~\ref{fig2ant} but a
linear scale for $f(\psi^{\prime})$.}
\end{figure}

In Eqs.~(\ref{eq6ant}) and (\ref{eq7ant}) the dimensionless
scaling variable $\psi '^2$ (in units of the Fermi energy) has the
physical meaning of the smallest kinetic energy which one of the
nucleons responding to an external probe can have~\cite{BCD+98} .
Since $\eta_F ^2\ll 1$ we write Eq.~(\ref{eq6ant}) as:
\begin{equation}\label{eq8ant}
f^\text{HSDFG}(\psi')\backsimeq\frac{3}{2}\int\limits_{|\psi'|}^{\infty}
x n(x)dx.
\end{equation}
It can be seen from Eq.~(\ref{eq8ant}) that, as expected, the
HSDFG system under consideration also ${\it exhibits}$ ${\it
superscaling}$. Eq.~(\ref{eq8ant}) has been obtained using the
approximations following~\cite{Sato}:
\begin{itemize}
    \item[i)] the Fermi momentum distribution of the initial
nucleon in the nucleus $\dfrac{3}{4\pi k_F^3}\theta(k_F-|k|)$ is
replaced by the MD (with dimension), namely $\int P(\vec{k},E)dE$,
where $P(\vec{k},E)$ is the spectral function, and
    \item[ii)] for $k> k_F$ the step function $\theta(k_F-|k|)$ is
retained to take into account approximately Pauli blocking for the
final nucleon.
\end{itemize}
In Figs.~\ref{fig2ant} and~\ref{fig3ant} we give the results for
the HSDFG scaling function (in logarithmic and linear scale,
respectively) calculated for different values of $k_F c$ from
$0.70$ to $0.28$ and compared to the result for the scaling
function in the RFG model. One can see that the HSDFG scaling
function is extended for large negative values of $\psi '$ in
contrast to the case of the RFG scaling function, but there is not
a good agreement with the experimental data. One can see in the
figures also the step behavior of the scaling function which
reflects the discontinuity of $n(k)$ at $k = k_F$. Due to these
reasons we consider in the second part of this Section the
relation between the asymptotic behavior of the MD and the $\psi
'$-scaling function.

\begin{figure}
\centering
\includegraphics[width=80mm]{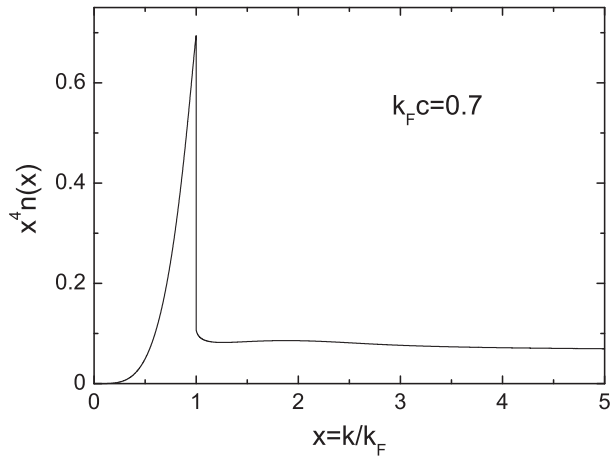}
\caption{\label{fig4ant} The momentum distribution in HSDFG $n(x)$
multiplied by $x^ 4 = (k/k_F)^4$.}
\end{figure}

A relationship between the NN force in nuclear medium
$\widetilde{V}_\text{NN}(k)$ and the asymptotic behavior of the
momentum distribution $n(k)$ was derived with great generality
in~\cite{Amado76a,Amado76b,Amado77}. In these articles Amado and
Woloshyn showed that the asymptotics of $n(k)$ at large values of
$k$ is a power-law decrease:
\begin{equation}\label{eq9ant}
n(k)\xrightarrow[k\rightarrow
\infty]{}\bigg[\frac{\widetilde{V}_\text{NN}(k)}{k^2}\bigg]^2,
\end{equation}
where ${\widetilde{V}}_\text{NN}(k)$ is the Fourier transform of
the NN interaction ${V}_\text{NN}(r)$. In the case of delta-forces
(\emph{i.e.} in the HSDFG) the asymptotics is $n(k) \sim
1/k^4$~\cite{Czyz}. It is still unknown if $k$ or $k/A$ must be
large for Eq.~(\ref{eq9ant}) to
apply~\cite{Amado76a,Amado76b,Amado77}. In principle, it is shown
in~~\cite{Sartor} that $n(k)$ in the HSDFG decreases faster than
$k^{-4}$, typically like $\sim 1/ k^{4+m}$, where $m>0$. In
Fig.~\ref{fig4ant} we show $x^4n(x)$ as a function of $x=k/k_F$.
It can be seen that the HSDFG $n(k)$ decreases approximately like
$\sim 1/ k^{4+m}$ with a small value of $m$, because the result
for $x^4n(x)$ is almost a straight line which decreases slowly
with the increase of $x \equiv k/k_F> 1$.

Further in this work we study the question about the general
feature of the NN force $\widetilde{V}_\text{NN}(k)$ that results
in an $n(k)$ with a power-law behavior that best agrees with the
scaling function. To this aim we assume an NN-potential
${V}_\text{NN}(r)$ different from a delta-function and calculate
the scaling function using different asymptotics for $n(k)$ in the
dilute Fermi gas at $k > k_F$. Therefore we look for the proper
value of $m$. For $k < k_F$ we use $n(k)$ [Eq.~(2)]
from~\cite{Sartor}, but for $k > k_F$ we use:
\begin{equation}\label{eq10ant}
n(k)=N\frac{1}{k^{4+m}}\quad\text{for}\quad k>k_F.
\end{equation}

The value of $N$ is obtained by the total normalization of $n(k)$
and is equal to
\begin{equation}\label{eq11ant}
N=\frac{0.24}{3}(1+m)k_F^{4+m}.
\end{equation}
The factor $0.24$ corresponds to the result for the part of the
normalization (for $k>k_F$) from the total normalization
condition:
\begin{equation}\label{eq12ant}
\frac{3}{4\pi k_F^3}\int n(\vec{k})d^3 \vec{k}=1.
\end{equation}

Finally, from Eq.~(\ref{eq8ant}) one can obtain the following
expression for the scaling function:
\begin{equation}\label{eq13ant}
f(\psi')=0.12\bigg(\frac{1+m}{2+m}\bigg)\frac{1}{|\psi'|^{2+m}}.
\end{equation}

\begin{figure}
\centering
\includegraphics[width=80mm]{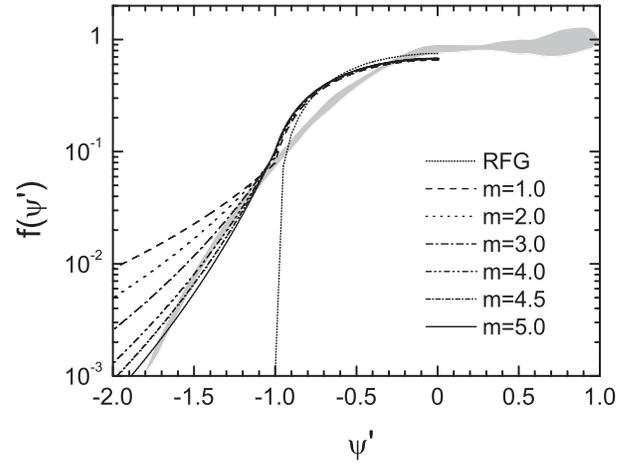}
\caption{\label{fig5ant} The scaling function in a dilute Fermi gas
calculated using Eq.~(\ref{eq13ant}) for different values of $m$ in
the asymptotics of the momentum distribution $n(k) \sim 1/k^{4+m}$
given in comparison with the RFG result. The grey area shows
experimental data taken from~\cite{DS99}.}
\end{figure}

In Fig.~\ref{fig5ant} we present the results for the scaling
function [Eq.~(\ref{eq8ant})] for different values of $m$,
compared with the RFG model result. One can see that agreement
with the experimental QE scaling function is achieved when the
value $m \approx  4.5$ is used in Eqs.~(\ref{eq10ant}),
(\ref{eq11ant}) and (\ref{eq13ant}). This means that the power-law
decrease of $n(k)$ which gives an optimal agreement with the data
is
\begin{equation}\label{eq14ant}
n(k)\approx\frac{1}{k^{8.5}}.
\end{equation}

We should note that this particular form of the power-law
asymptotics is close to that obtained in the CDFM~\cite{AHP88}
\begin{equation}\label{eq15ant}
n(k) \sim \frac{1}{k^8},
\end{equation}
i.e. it corresponds to $n(k)\sim 1/ k^{4+m}$ with $m=4$. The
inverse Fourier transform of $\widetilde{V}_\text{NN}(k)$ for $m =
4$ and $m = 5$ gives ${V}_\text{NN}(r)\sim 1/r$ and
${V}_\text{NN}(r)\sim 1/r^{1/2}$, respectively.

We would like to emphasize the consistency of both the optimal
asymptotics of $n(k)$ for the dilute Fermi gas found in this work
[Eq.~(\ref{eq14ant})] with that in the CDFM [Eq.~(\ref{eq15ant})].
As was shown in~\cite{AGK+04,AGK+05,AIG+06,AIG+06a} the calculated
QE scaling function $f(\psi ')$ in the CDFM agrees well with the
experimental scaling function. This fact shows that the behavior
of the QE scaling function depends mainly on the particular form
of the power-law asymptotics of the nucleon momentum distribution.
This is proved on our work by the similarities of the result for
the case of an interacting dilute Fermi gas with that one obtained
in the CDFM as a model accounting for NN correlations in realistic
finite nuclear systems.

\section[]{CONCLUSIONS\label{sect3ant}}

The results of the present work can be summarized as follows:
\begin{description}
  \item[i)] The superscaling observed in inclusive electron scattering from
nuclei is considered within the model of dilute Fermi gas with
interactions between particles. The latter gives an improvement in
comparison with the results of the relativistic non-interacting
Fermi gas model allowing one to describe the QE scaling function for
$\psi ' < -1$, whereas the RFG model gives $f(\psi ') = 0$ in this
region.
  \item[ii)] It is established that the hard-sphere (with delta-forces between
nucleons) approximation for the dilute Fermi gas is quite a rough
one. The use of more realistic NN forces leading to $m  \simeq 4.5$
instead of $m = 0$ (for delta-force) in the well-known power-law
asymptotics of the momentum distribution $n(k) \sim  1/k^{4+m}$ at
large $k$ leads to a good explanation of the data for the
$\psi'$-scaling function in inclusive electron scattering from a
wide range of nuclei.
  \item[iii)] The asymptotics of $n(k) \sim  1/k^{8.5}$ found in the dilute Fermi
gas by optimal fit to the data for $f(\psi ')$ is similar to that
in the CDFM ($\sim  1/k^8$)~\cite{AHP88} which, being a
theoretical correlation model, describes the superscaling in the
quasielastic part of the electron-nucleus scattering. Thus, the
momentum distribution in the dilute Fermi-gas model with realistic
NN forces can serve as an ``effective'' momentum distribution (a
step-like one with a discontinuity) which gives a similar result
for $f(\psi ')$ as the correlation methods for realistic finite
nuclear systems. It can be concluded that the momentum
distribution with asymptotics from $\sim  1/k^8$ to $\sim
1/k^{8.5}$ is the proper one to explain the phenomenological shape
of the scaling function obtained from inclusive QE electron
scattering.
\end{description}

As already mentioned, the superscaling is due to the {\it specific
high-momentum tail} of $n(k)$ similar for all nuclei which is
known to be caused by the short-range and tensor correlations
related {\it to peculiarities of the NN forces} near their core.
The main result of the present work might be the observation that
the values of $f(\psi ')$ for $\psi ' < -1$ depend on {\it the
particular form} of the power-law asymptotics of $n(k)$ at large
$k$ which is related to {\it a corresponding particular behavior}
of the {\it in-medium NN forces} around the core. Namely, we point
out that the power-law decrease of $n(k)$ as $\sim 1/k^{4+m}$ with
$m \simeq 4.5$ in the interacting dilute Fermi gas is the proper
one and it is close to that obtained in CDFM ($m=4$ \cite{AHP88})
which describes the superscaling correctly as well. The NN force
for $m=4$ is expected to go as ${V}_\text{NN}(r)\sim 1/r$ and for
$m=5$ to go as ${V}_\text{NN}(r)\sim (1/r)^{1/2}$. Hence, the
present study allows one to conclude that the important property
of the repulsive short-range core (leading to NN correlations and
high-momentum tail of $n(k)$) is that it goes to infinity for $r
\rightarrow 0$ as $1/r$ or softer. The link between the asymptotic
behavior of $n(k)$ and NN force implies that inclusive QE electron
scattering from nuclei provides important information on the NN
forces in the nuclear medium.

\begin{acknowledgments}
This work was partly supported by the Bulgarian Science Fund under
Contracts No.~$\Phi$-1416 and $\Phi$-1501 and by the Ministerio de
Educaci\'{o}n y Ciencia (Spain) under Contract FIS2005-00640. One
of the authors (M.K. Gaidarov) is grateful for warm hospitality to
the CSIC and for support during his stay there to the State
Secretariat of Education and Universities of Spain (N/Ref.
SAB2005--0012).

\end{acknowledgments}


\begin{thebibliography}{99}

\bibitem{West75}
G. B. West, Phys. Rep. {\bf 18}, 263 (1975).

\bibitem{Sick80} I. Sick, D. B. Day, and J. S. McCarthy, Phys.
Rev. Lett. {\bf 45}, 871 (1980).

\bibitem{Cio87}
C. Ciofi degli Atti, E. Pace, and G. Salm\`{e}, Phys. Rev. C {\bf
36}, 1208 (1987).

\bibitem{Day90} D. B. Day, J. S. McCarthy, T. W.
Donnelly, and I. Sick, Annu. Rev. Nucl. Part. Sci. {\bf 40}, 357
(1990).

\bibitem{Cio90}
C. Ciofi degli Atti, E. Pace, and G. Salm\`{e}, Phys. Rev. C {\bf
43}, 1155 (1991).

\bibitem{Cio96}
C. Ciofi degli Atti and S. Simula, Phys. Rev. C {\bf 53}, 1689
(1996).

\bibitem{CW97} C. Ciofi degli Atti and G. B. West, nucl-th/9702009.

\bibitem{CW99} C. Ciofi degli Atti and G. B. West, Phys. Lett. B \textbf{458}, 447 (1999).

\bibitem{FCW00} D. Faralli, C. Ciofi degli Atti, and G. B. West, in
\emph{Proceedings of 2nd International Conference on Perspectives in
Hadronic Physics}, ICTP, Trieste, Italy, 1999, edited by S. Boffi,
C. Ciofi degli Atti, and M. M. Giannini (World Scientific,
Singapore, 2000), p.~75.

\bibitem{AMD+88} W. M. Alberico, A. Molinari, T. W. Donnelly, E. L.
Kronenberg, and J. W. Van Orden, Phys. Rev. C \textbf{38}, 1801
(1988).

\bibitem{BCD+98} M. B. Barbaro, R. Cenni, A. De Pace, T. W. Donnelly,
and A. Molinari, Nucl. Phys. A \textbf{643}, 137 (1998).

\bibitem{DS99l} T. W. Donnelly and I. Sick, Phys. Rev. Lett.
\textbf{82}, 3212 (1999).

\bibitem{DS99} T. W. Donnelly and I. Sick, Phys. Rev. C \textbf{60},
065502 (1999).

\bibitem{MDS02} C. Maieron, T. W. Donnelly, and I. Sick, Phys. Rev. C
\textbf{65}, 025502 (2002).

\bibitem{AGK+04} A. N. Antonov, M. K. Gaidarov, D. N. Kadrev, M. V.
Ivanov, E. Moya de Guerra, and J. M. Udias, Phys. Rev. C
\textbf{69}, 044321 (2004).

\bibitem{AGK+05} A. N. Antonov, M. K. Gaidarov, M. V.
Ivanov, D. N. Kadrev, E. Moya de Guerra, P. Sarriguren, and J. M.
Udias, Phys. Rev. C \textbf{71}, 014317 (2005).

\bibitem{AIG+06} A. N. Antonov, M. V. Ivanov, M. K. Gaidarov, E. Moya de Guerra,
P. Sarriguren, and J. M. Udias, Phys. Rev. C \textbf{73}, 047302;
\textbf{73}, 059901 (E) (2006).

\bibitem{AIG+06a} A. N. Antonov, M. V. Ivanov, M. K. Gaidarov, E. Moya de Guerra,
J. A. Caballero, M. B. Barbaro, J. M. Udias, and P. Sarriguren,
Phys. Rev. C \textbf{74}, 054603 (2006).

\bibitem{Benhar2006} O. Benhar, D. Day, and I. Sick, nucl-ex/0603029.

\bibitem{BCD+04} M. B. Barbaro, J. A. Caballero, T. W. Donnelly, and
C. Maieron, Phys. Rev. C \textbf{69}, 035502 (2004).

\bibitem{AlvRuso03} L. Alvarez-Ruso, M. B. Barbaro, T. W. Donnelly, and
A. Molinari, Nucl. Phys. A \textbf{724}, 157 (2003).

\bibitem{Amaro2002} J. E. Amaro, M. B. Barbaro, J. A. Caballero, T. W.
Donnelly, and A. Molinari, Nucl. Phys. A \textbf{697}, 388 (2002);
\textbf{723}, 181 (2003); Phys. Rept. {\bf 368}, 317 (2002).

\bibitem{Pace03} A. De Pace, M. Nardi, W. M. Alberico, T. W. Donnelly,
and A. Molinari, Nucl. Phys. A \textbf{726}, 303 (2003); {\it ibid.}
\textbf{741}, 249 (2004).

\bibitem{ANP79+} A.~N.~Antonov, V.~A.~Nikolaev, and I.~Zh.~Petkov,
Bulg. J. Phys. \textbf{6}, 151 (1979); Z. Phys. A \textbf{297}, 257
(1980); \textit{ibid.} \textbf{304}, 239 (1982); Nuovo Cimento A
\textbf{86}, 23 (1985); Nuovo Cimento A \textbf{102}, 1701 (1989);
A. N. Antonov \emph{et al.}, Phys. Rev. C \textbf{50}, 164 (1994).

\bibitem{AHP88} A.~N.~Antonov, P.~E.~Hodgson, and I.~Zh.~Petkov,
\textit{Nucleon Momentum and Density Distributions in Nuclei}
(Clarendon Press, Oxford, 1988); \textit{Nucleon Correlations in
Nuclei} (Springer-Verlag, Berlin-Heidelberg-New York, 1993).

\bibitem{AGI+02} A. N. Antonov, M. K. Gaidarov, M. V. Ivanov, D. N.
Kadrev, G. Z. Krumova, P. E. Hodgson, and H. V. von Geramb, Phys.
Rev. C \textbf{65}, 024306 (2002).

\bibitem{CK95} J. Carbonell and V. A. Karmanov, Nucl. Phys. A
\textbf{581}, 625 (1995).

\bibitem{CDK98} J. Carbonell, B. Desplanques, V. A. Karmanov, and
J.-F. Mathiot, Phys. Rep. \textbf{300}, 215 (1998).

\bibitem{Amaro2005} J. E. Amaro, M. B. Barbaro, J. A. Caballero, T. W. Donnelly,
A. Molinari, and I. Sick, Phys. Rev. C \textbf{71}, 015501 (2005).

\bibitem{Caballero2005} J. A. Caballero, J.~E.~Amaro, M. B. Barbaro,
T. W. Donnelly, C. Maieron, and J. M. Udias, Phys. Rev. Lett.
\textbf{95}, 252502 (2005).

\bibitem{Caballero2006} J.A. Caballero, Phys. Rev. C \textbf{74},
015502 (2006).

\bibitem{Meucci20041} A. Meucci, C. Giusti, and F. D. Pacati,
Nucl. Phys. A \textbf{744}, 307 (2004).

\bibitem{Amaro2006} J. E. Amaro, M. B. Barbaro, J.A. Caballero,
and T.W. Donnelly, Phys. Rev. C \textbf{73}, 035503 (2006).

\bibitem{Barbaro2006} M. B. Barbaro, Nucl. Phys. B, Proc. Suppl. \textbf{159}, 186 (2006); nucl-th/0602011.

\bibitem{Martinez2006} M. C. Martinez, P. Lava, N. Jachowicz, J.
Ryckebusch, and J. M. Udias, Phys. Rev. C \textbf{73}, 024607
(2006).

\bibitem{Nieves2005} J. Nieves, M. Valverde, and M. J.
Vicente-Vacas, Nucl. Phys. B, Proc. Suppl. \textbf{155}, 263 (2006);
nucl-th/0510010.

\bibitem{Barbaro2005} M. B. Barbaro, J. E. Amaro, J. A. Caballero,
T. W. Donnelly, and A. Molinari, Nucl. Phys. B, Proc. Suppl.
\textbf{155}, 257 (2006); nucl-th/0509022.

\bibitem{Maieron2003} C. Maieron, M. C. Martinez, J. A. Caballero, and
J. M. Udias, Phys. Rev. C \textbf{68}, 048501 (2003).

\bibitem{Meucci20042} A. Meucci, C. Giusti, and F. D. Pacati,
Nucl. Phys. A \textbf{739}, 277 (2004); Nucl. Phys. A \textbf{773},
250 (2006).

\bibitem{Benhar} O. Benhar, Nucl. Phys. B, Proc. Suppl. \textbf{139}, 15 (2005); nucl-th/0408045;
 O. Benhar and N. Farina, Nucl. Phys. B, Proc. Suppl. \textbf{139}, 230 (2005); nucl-th/0407106;
 O. Benhar, N. Farina, H. Nakamura, M. Sakuda, and R. Seki, Nucl. Phys. B, Proc. Suppl. \textbf{155}, 254 (2006);
 hep-ph/0510259; Phys. Rev. D \textbf{72}, 053005 (2005).

\bibitem{Co} G. Co', Nucl. Phys. B, Proc. Suppl. \textbf{159}, 192 (2006);
nucl-th/0601034; A. Botrugno and G. Co', Nucl. Phys. A
\textbf{761}, 200 (2005); M. Martini, G. Co', M. Anguiano, and A.
M. Lallena, nucl-th/0701031.

\bibitem{Leither} T. Leitner, L. Alvarez-Ruso, and U. Mosel, Phys. Rev. C \textbf{73}, 065502 (2006).

\bibitem{Sato} B. Szczerbinska, T. Sato, K. Kubodera, and T.-S. H. Lee, nucl-th/0610093.

\bibitem{Sartor} R. Sartor and C. Mahaux, Phys. Rev. C \textbf{21},
1546 (1980); Phys. Rev. C \textbf{25}, 677 (1982).

\bibitem{Czyz} W. Czy\.{z} and K. Gottfried, Nucl. Phys. {\bf 21}, 676
(1961).

\bibitem{Migdal} A. B. Migdal, Zh. Eksp. Teor. Fiz. \textbf{32}, 333
(1957).

\newblock [Sov. Phys. JETP \textbf{5}, 333 (1957)].

\bibitem{Galitskii} V. M. Galitskii, Zh. Eksp. Teor. Fiz. \textbf{34}, 151
(1958).

\newblock [Sov. Phys. JETP \textbf{7}, 104 (1958)].

\bibitem{Belyakov} V. A. Belyakov, Zh. Eksp. Teor. Fiz. \textbf{40}, 1210
(1961).

\newblock [Sov. Phys. JETP \textbf{13}, 850 (1961)].


\bibitem{Amado76a} R. D. Amado and R. M. Woloshyn, Phys. Lett. B \textbf{62}, 253
(1976).

\bibitem{Amado76b} R. D. Amado, Phys. Rev. C \textbf{14}, 1264
(1976).

\bibitem{Amado77} R. D. Amado and R. M. Woloshyn, Phys. Rev. C \textbf{15}, 2200
(1977).

\end{thebibliography}
\end{document}